\documentclass[10pt, preprint2]{aastex}

\usepackage{epsfig}
\usepackage{natbib}
\def\plotone#1{\centering \leavevmode
\includegraphics[clip=, width=.75\columnwidth]{#1}}

\newcommand{\R}[0]{\mathcal{R}}
\newcommand{\pn}[1]{\mbox{$(#1)$}}
\newcommand{\spa}{\mbox{ }}
\def\gsim{\;\rlap{\lower 2.5pt
 \hbox{$\sim$}}\raise 1.5pt\hbox{$>$}\;}
\def\lsim{\;\rlap{\lower 2.5pt
   \hbox{$\sim$}}\raise 1.5pt\hbox{$<$}\;}

\begin{document}


\title{A Possible Dearth of Hot Gas in Galaxy Groups at Intermediate Redshift}

\author{David S. Spiegel\altaffilmark{1}, Frits Paerels\altaffilmark{1},
Caleb A. Scharf\altaffilmark{1}}

\affil{$^1$Department of Astronomy, Columbia University, 550 West 120th Street, New York, NY 10027}

\vspace{0.5\baselineskip}

\email{dave@astro.columbia.edu, frits@astro.columbia.edu,
caleb@astro.columbia.edu}

\begin{abstract}
We examine the X--ray luminosity of galaxy groups in the \textsl{CNOC2} survey,
at redshifts $0.1 < z < 0.6$.  Previous work examining the gravitational
lensing signal of the \textsl{CNOC2} groups has shown that they are likely to be
genuine, gravitationally bound objects.  Of the 21 groups in the field of view
of the \textsl{EPIC--PN} camera on \textsl{XMM--Newton}, not one was visible
in over 100~ksec of observation, even though three of the them have velocity
dispersions high enough that they would easily be visible if their luminosities
scaled with their velocity dispersions in the same way as nearby groups'
luminosities scale.  We consider the possibility that this is due to
the reported velocity dispersions being erroneously high, and conclude that
this is unlikely.  We therefore find tentative evidence that groups at
intermediate redshift are underluminous relative to their local cousins.
\end{abstract}

\keywords{cosmology: observations -- galaxies: clusters: general --
X--rays: galaxies: clusters}

\section{Introduction}
\label{sec:intro}
According to the paradigm of hierarchical structure formation, 
as matter falls from low--density environments to high--density environments
(i.e., clusters of galaxies) it passes through stages of intermediate density,
namely, groups.  Rich clusters of galaxies, containing as many as a thousand
member--galaxies or more, are visually quite prominent and have therefore
attracted research interest for over 70 years.  Over the last several decades,
however, it has become increasingly apparent that small groups of galaxies,
containing fewer than 50 members and often containing as few as 3--10,
constitute by far the most common environment in which galaxies are found in
the universe today, and are therefore the dominant stage of
structure--evolution at the present epoch.  In order to understand the
structure and evolution of matter in the universe, then, we must understand the
properties of groups of galaxies.

In rich clusters, a large fraction of the baryonic mass ($\gtrsim 80\%$) exists
as diffuse, hot, intracluster gas -- see \citet{ettori_et_al2003} and the review
by \citet{rosati_et_al2002} and references therein -- that has
been detected with space--based X--ray telescopes for over 30 years
\citep{giacconi_et_al1974, rowanrobinson+fabian1975, schwartz1978}.
If small galaxy groups have a similar hot--gas component, then a large fraction
of the baryonic mass of the universe could be hiding in these groups
\citep[and references therein]{fukugita+peebles2004}.
Alternatively, if groups do not contain a diffuse, extended intragroup medium
(IGM), this would be troubling news for the idea that structure forms
in the hierarchical fashion that is widely assumed.

In the last 15 years, the improved resolution and sensitivity of new X--ray
telescopes have allowed us to begin to study the X--ray properties of groups
of galaxies.  \textsl{ROSAT} observations indicate that in the local universe,
half to three quarters of groups have a hot, X--ray emitting IGM
\citep{mulchaey2000}.  
\citet*[hereafter PT04]{plionis+tovmassian2004} examine the relationship
between the X--ray luminosity ($L_X$) and the velocity dispersion
($\sigma_v$) of nearby groups in the \citet[hereafter M03]{mulchaey_et_al2003}
catalog (although there remain significant uncertainties associated with the
analysis).
At higher redshift, however, a systematic study of this kind has not yet been
performed, although some recent evidence suggests that the relationship at
$z \gsim 0.5$ is different from the local one: a deep \textsl{Chandra}
observation failed to reveal X--ray emission from groups discovered by the
\textsl{DEEP2} Galaxy Redshift Survey, which \citet{fang_et_al2006astroph}
find strongly suggests that the groups in their survey (at $z \sim 1$) are
less X--ray luminous than the nearby relationship would predict.

A recent deep imaging and spectroscopic study by the Canadian Network for
Observational Cosmology Field Galaxy Redshift Survey (\textsl{CNOC2}) made
it possible to identify a large number of galaxy groups at intermediate
redshift ($0.1 < z < 0.6$) \citep{carlberg2001eg, carlberg2001gg}.
Gravitational lensing analysis of the \textsl{CNOC2}
fields yields statistical lensing masses of these groups
\citep{hoekstra_et_al2001}.
This data set, therefore, provides an ideal laboratory for
studying the X--ray properties of groups at these redshifts.

Fortunately, the data archive for \textsl{XMM--Newton} contains
approximately 110 ksec worth of observations mostly overlapping one of the
\textsl{CNOC2} fields (21 of the groups are in the field of view of the
\textsl{EPIC--PN} camera; see Table~\ref{ta:21Groups}).
If the relation between $L_X$ and $\sigma_v$ of groups at these redshifts
is the same as the relation that holds for nearby groups,
then we will argue that a few of the most
massive groups ought to have been visible in the aggregate data from
these observations.  In this analysis, we looked for whether the X--ray
photons received are at all correlated to the groups:
\pn{i} Are there more (or fewer) photons
where there are groups than where there are not?
\pn{ii} Is the spectral energy distribution of the
photons where there are groups different from that of the background? 

In \S~\ref{sec:obs}, we describe the data we used and how we reduced it.  In
\S~\ref{sec:spatial} and \S~\ref{sec:spec}, we describe our analysis of the
data -- spatial analysis and spectral analysis, respectively.
We found no
spatial or spectral evidence for the groups; and in \S~\ref{sec:disc}, we
discuss what we had expected the data to look like, and how surprising it is
that we failed to detect the groups.  In \S~\ref{sec:conc}, we attempt to draw
conclusions, and we speculate as to the reasons why we did not detect the
groups.  Finally, in Appendix~I, we give a brief analysis of several
different methods of fitting lines to data; and in Appendix~II, we describe
how we estimate the probability that groups at intermediate redshift share the
same X--ray luminosity function as those in the local universe.

\section{Observations and Data Reduction}
\label{sec:obs}
The \textsl{CNOC2} survey comprised four fields on the sky.  We examined
the X--ray properties of the optically identified galaxy groups in the
1447+09 field \citep{carlberg2001eg, carlberg2001gg}.
For our analysis, we used the \textsl{EPIC--PN} data from
three publicly available observations of that field in the
\textsl{XMM--Newton} data archive, of duration 33 ksec, 33 ksec, and 43 ksec.
We did not use the corresponding \textsl{EPIC--MOS} data in the current
analysis because of the lower sensitivity of the \textsl{MOS} to low energy
photons, and the attendant modest increase in sensitivity in
background--limited images: including \textsl{MOS} would have increased the
number of counts at the locations of the groups
in our analysis by a factor of only $\sim 50\%$.  The relevant summary data
on the 21 groups in the field of view of \textsl{EPIC--PN}, from
\citet{carlberg2001eg}, is presented in Table~\ref{ta:21Groups}.

For basic data reduction, including generating good--time--intervals (GTI)
files to deal with periods of high solar activity, removing known
hot--pixels, and generating images and exposure maps for energy bands of
interest, we used \textsl{XMM--Newton} Science Analysis Software (SAS)
release 6.1.0.  The light--curves used to generate GTI files consisted of
integrated flux in the range 0.3--15.0~keV across the whole detector.  Time
intervals of particularly high background ({\tt RATE>9}) were flagged as
periods of flaring and were removed from the data.  For the purpose of the
analysis described below, we considered four energy bands:
0.3--0.8~keV, 0.8--1.5~keV, 1.5--4.5~keV, and 0.2--1.0~keV, which we will
hereafter refer to as bands A, B, C, and D, respectively.  To make
images in each band, we selected events in the GTI with {\tt PATTERN<=4} and
{\tt FLAG==0}.  Images and corresponding exposure maps were made at $2''$ per
pixel, roughly twice the resolution of the intrinsic pixel size of
\textsl{EPIC--PN} ($4.1''$). 

Since we were looking for faint emission (from extended sources) that was
expected to outshine the background by only a slim margin, it was essential to
remove point sources from the images.
In addition, it was crucial to characterize the background
carefully before further analysis so as to maximize the accuracy of the
measurements of both the background flux level and the group luminosities. 
Two SAS tasks -- {\tt eboxdetect} and {\tt emldetect} -- were used to find and
remove point sources in all observations, in all energy bands; locations of
point sources were then masked in the images and exposure maps.  In order to
characterize the background, we used two methods that are described in
detail in the next section.  In short, one was to smooth the point--source
excluded image, and the other was to sample the average count--rate over
the groups--excluded image.

\begin{deluxetable}{rcccccc}
\tablecolumns{7}
\small
\tablecaption{The 21 \textsl{CNOC2} Groups in the Field of View}
\tablehead{\colhead{Group ID} & \colhead{R.A. (J2000)} & \colhead{Decl. (J2000)} & \colhead{$z$} & \colhead{$N_z$} & \colhead{$\sigma_v$ ($\rm km~s^{-1}$)} & \colhead{Estimated Mass ($h^{-1} M_\sun$)}}
\startdata
 1 & 14 49 42.453  & 09 02 44.76  & 0.165  & 3  & 164 $\pm$ 126 & $1.3 \times 10^{13}$ \\
 2 & 14 48 55.799  & 09 08 48.48  & 0.229  & 3  & 162 $\pm$ 139 & $7.2 \times 10^{12}$ \\
 3 & 14 48 57.854  & 08 57 05.58  & 0.262  & 4  & 229 $\pm$  77 & $1.2 \times 10^{13}$ \\
 4 & 14 49 28.406  & 08 51 54.07  & 0.270  & 3  & 104 $\pm$  92 & $3.0 \times 10^{12}$ \\
 5 & 14 49 13.637  & 08 50 10.34  & 0.271  & 4  & 112 $\pm$  66 & $3.0 \times 10^{12}$ \\
 6 & 14 49 44.220  & 08 57 37.84  & 0.273  & 3  & 165 $\pm$ 120 & $8.1 \times 10^{12}$ \\
 7 & 14 48 49.049  & 08 57 50.15  & 0.306  & 3  &  93 $\pm$  65 & $1.3 \times 10^{12}$ \\
 8 & 14 50 20.387  & 09 06 00.62  & 0.306  & 4  & 199 $\pm$ 161 & $1.6 \times 10^{13}$ \\
 9 & 14 50 13.703  & 08 57 37.28  & 0.325  & 4  & 175 $\pm$ 151 & $9.9 \times 10^{12}$ \\
10 & 14 49 03.618  & 09 07 02.86  & 0.359  & 4  &  82 $\pm$  66 & $1.6 \times 10^{12}$ \\
11 & 14 49 40.860  & 09 02 15.68  & 0.372  & 3  & 126 $\pm$  94 & $2.8 \times 10^{12}$ \\
12 & 14 50 22.850  & 09 01 14.74  & 0.373  & 4  &  44 $\pm$  41 & $2.1 \times 10^{11}$ \\
13 & 14 49 49.201  & 08 51 23.08  & 0.374  & 4  & 291 $\pm$ 193 & $3.4 \times 10^{13}$ \\
14 & 14 50 00.825  & 08 49 06.65  & 0.394  & 4  & 308 $\pm$ 257 & $3.6 \times 10^{13}$ \\
15 & 14 49 14.155  & 09 11 15.38  & 0.394  & 3  & 394 $\pm$ 406 & $4.8 \times 10^{13}$ \\
16 & 14 49 55.433  & 08 56 11.39  & 0.394  & 3  & 507 $\pm$ 469 & $7.8 \times 10^{13}$ \\
17 & 14 49 32.746  & 09 03 41.28  & 0.468  & 4  & 488 $\pm$ 417 & $7.5 \times 10^{13}$ \\
18 & 14 49 29.974  & 09 09 08.15  & 0.469  & 4  & 217 $\pm$ 224 & $8.9 \times 10^{12}$ \\
19 & 14 49 31.428  & 09 05 05.22  & 0.472  & 3  & 123 $\pm$  99 & $4.9 \times 10^{12}$ \\
20 & 14 49 23.516  & 08 58 47.83  & 0.511  & 3  & 565 $\pm$ 668 & $1.2 \times 10^{14}$ \\
21 & 14 49 23.763  & 08 55 17.84  & 0.543  & 3  & 151 $\pm$ 132 & $8.7 \times 10^{12}$ \\
\enddata
\label{ta:21Groups}
\vspace{-0.4cm}
\tablenotetext{a}{ID Numbers are assigned to the groups that were used for
the analysis in this paper.  Groups are listed in order of increasing redshift
($z$).  $N_z$ is the number of galaxies with identified redshifts in each
group.  $\sigma_v$ is line-of-sight velocity dispersion of group galaxies.
Estimated mass, and all other numbers in this table (except ID Number), are
taken from \citet{carlberg2001eg}.
}
\end{deluxetable}

\section{Spatial Analysis}
\label{sec:spatial}
We investigated whether the spatial locations of received photons were at
all correlated with the groups.  This inquiry required a metric of surface
brightness that would allow us to compare quantitatively different patches
of sky.  The idea behind the desired metric is to find the average number
of counts per unit sky area per unit time within the patch.  A reasonable
first guess of the metric would therefore be the sum of the pixel values in
the patch,
divided by the number of pixels, divided by the total exposure time.  In
an experiment in which the pixels have nonuniform effective exposure time,
however, they have unequal sensitivity.  Maximizing the sensitivity of
such an experiment requires weighting more heavily those pixels that are
more sensitive in the observation.  A reasonable second guess of the metric
of the surface brightness of a patch of $N$ pixels, then, is
\[
\R \equiv k \times \frac{\sum_{i=1}^N(p_i/e_i)w_i}{\sum_{i=1}^Nw_i},
\]
where $k$ is the number of square arcseconds per pixel (in our images,
$k = 4 {\rm~arcsec^2~pixel^{-1}}$), $p_i$ is the number of counts in the $i$th
pixel of the patch, $e_i$ is the effective exposure time seen by
that pixel, and $w_i$ is an arbitrary weighting factor.  In order to maximize
the signal--to--noise ratio of $\R$, the appropriate weighting factor is
$w_i = e_i$, so in the end the metric reduces to
\begin{equation}
\label{eq:metric}
\R \equiv k \times \frac{\sum_{i=1}^Np_i}{\sum_{i=1}^Ne_i}.
\end{equation}

In order to determine the average surface brightness of the groups, we had
to know which patches of sky to define as the groups.  For each group, we
defined a circular aperture within which to search for X--ray emission.  
The best way to assign radii to groups is open to debate,
so we tried various ways, including two methods that depend on the
groups' properties -- $r_{500}$ and $r_{200}$,
calculated with formulae from the literature -- and several that do not.
The dependence
of $r_{500}$ \citep{osmond+ponman2004} and $r_{200}$ \citep{mahdavi+geller2004}
on redshift $z$ and line--of--sight velocity dispersion $\sigma_v$ are given by
\begin{eqnarray}
\label{eq:r500}
r_{500} & = & 130 {\rm~kpc} \left( \frac{\sigma_v}{100 {\rm~km~s^{-1}}} \right) \left( \frac{H(z)}{73 {\rm~km~s^{-1}~Mpc^{-1}}} \right)^{-1} \\
\label{eq:r200}
r_{200} & = & 230 {\rm~kpc} \left( \frac{\sigma_v}{100 {\rm~km~s^{-1}}} \right) \left( \frac{H(z)}{73 {\rm~km~s^{-1}~Mpc^{-1}}} \right)^{-1},
\end{eqnarray}
where $H(z)$ is the value of the Hubble constant at redshift $z$.  All
calculations in this paper assume $H_0 = 73 {\rm~km~s^{-1}~Mpc^{-1}}$,
$\Omega_M = 0.3$, and $\Omega_\Lambda = 0.7$.  We also tried assigning a
fixed physical size to each group ($125 {\rm~kpc}$, $250 {\rm~kpc}$, and
$500 {\rm~kpc}$).  In each case, physical size was converted to angular
size using $\alpha = r/D_A(z)$, where $r$ is the group's physical size and
$D_A(z)$ is its angular diameter distance.
Happily, our main scientific results were independent of how we calculated
the size of the aperture around each group -- and this is an important point,
on which we will elaborate below.  In this paper, we will present
results for $250 {\rm~kpc}$ apertures around each group.

The $\R$--value of a group
(according to equation~(\ref{eq:metric}), within the set of pixels defined
by the group's angular radius $\alpha$)
cannot by itself be converted to the
group's luminosity, because it includes counts from the background
that are unrelated to the groups.  The background flux level must
therefore be calculated and subtracted.  The true measure of the
average surface brightness of each group is the difference between the
group's $\R$--value and the $\R$--value of the background: 
\begin{equation}
\label{eq:brightness}
SB_g \equiv \R_g - \R_{BG}.
\end{equation}
We calculated the average surface brightness of the background in two
different ways, and we also calculated the local background corresponding to
each group, which gave us multiple measures of the surface brightness $SB_g$
of each group.

We calculated the average background for the whole image in two ways. 
First, for each energy band, we determined the $\R$--value of the image
for all pixels not contained in any group's aperture.
Second, we created
a background image for each energy band by smoothing that band's image
with a Gaussian smoothing radius of 60 pixels and, as before, determined
the $\R$--value of the background image for all pixels not contained in any
group's aperture.  These two methods of determining the average
background of the whole image 
yielded nearly identical results.  According to \citet{scharf2002},
at redshift of $\sim 0.5$,
low mass objects such as groups should be seen most readily in a
low--energy band such as D or A. 
Since D--band contains more counts than A--band, much of our analysis
deals with D--band data.  The different measurements of the whole--detector
background were similar; in D--band, for example, the background
rate was measured to be $3.26$ or
$3.08\times 10^{-6}{\rm~cts~arcsec^{-2}~s^{-1}}$, depending on 
whether the image or the smoothed (background) image was used.

We also measured the local background corresponding to each group by
taking an annulus of inner radius $1.05\alpha$ and outer radius
$1.20\alpha$ around each group, and finding the $\R$--value of the
background image in the annulus surrounding the group.
These local background measurements ranged from $2.44$ to
$4.21\times 10^{-6}{\rm~cts~arcsec^{-2}~s^{-1}}$, and had a median of
$2.77\times 10^{-6}{\rm~cts~arcsec^{-2}~s^{-1}}$ and a mean of
$2.96\times 10^{-6}{\rm~cts~arcsec^{-2}~s^{-1}}$.

\begin{figure}[t]
\plotone{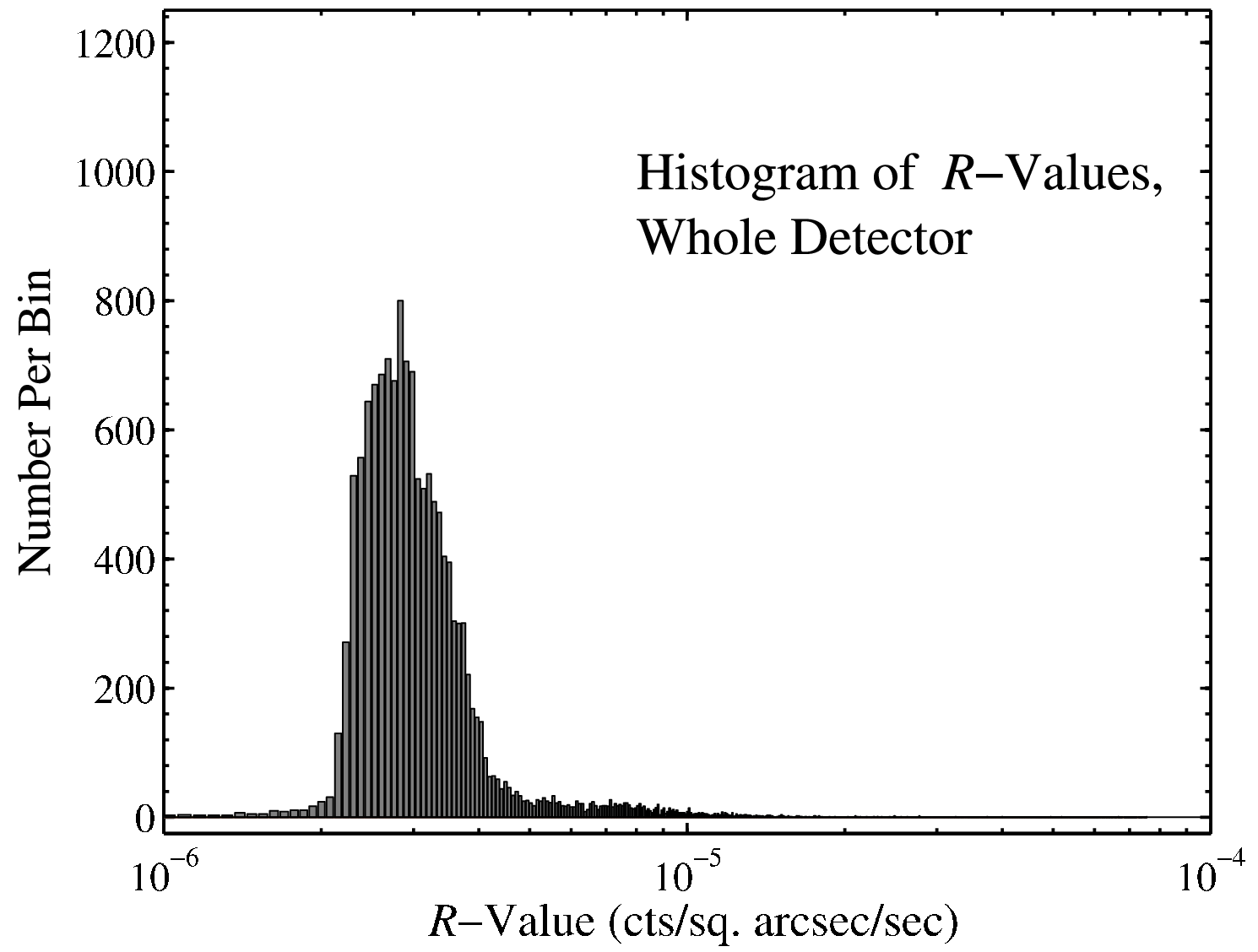}
\caption{Histogram of the result of 
20,000 random samples of the spatial variability of the D--band background.
The number of samples in each $\R$--bin is plotted.  The bin--size is
$2\times 10^{-7}$ counts per square arcsecond per second.  The mode, median,
and mean values in this histogram are consistent with other measures of the
background.
}
\label{fig:histfig}
\end{figure}

Finally, to understand the spatial variability of the background, in each
energy band we placed an aperture of radius $0.9\arcmin$ (27 pixels) in
20,000 random locations on the point--source--excluded image, and
measured the $\R$--value within each aperture.  This procedure gives an
empirical measure of the probability distribution function
that describes the surface--brightness in apertures of approximately the sizes
of the groups.  Results for D--band are displayed in the histogram in
Figure~\ref{fig:histfig}.  For D--band, the modal $\R$--value of the samples
of the background is $2.8\times10^{-6}{\rm~cts~arcsec^{-2}~s^{-1}}$;
the median $\R$--value, $3.0\times10^{-6}{\rm~cts~arcsec^{-2}~s^{-1}}$,
is similar; and the mean,
$3.5\times10^{-6}{\rm~cts~arcsec^{-2}~s^{-1}}$, is somewhat higher 
owing to a few anomalously high measurements that resulted from apertures
with small numbers 
of pixels near the edge of the detector or near masked regions. 

Surprisingly, the $\R$--values of the groups were frequently less than
those of the background.  This means both that $\R$--values of individual groups
were often less than the corresponding $\R$--values of the local background, and
that the stacked $\R$--value for all 21 groups,
$2.78\times 10^{-6}{\rm~cts~arcsec^{-2}~s^{-1}}$, was $10\%$ {\it less} than
the lesser of the two measures of the average background.  In fact, the
average value of the surface brightness of groups (calculated with 
equation~(\ref{eq:brightness}) and using the local background $\R$--value for
$\R_{BG}$; average was weighted by the product of the number of pixels and the
effective exposure time) turned out to be negative:
$-8.6\times 10^{-8}{\rm~cts~arcsec^{-2}~s^{-1}}$.

\begin{figure}[t]
\plotone{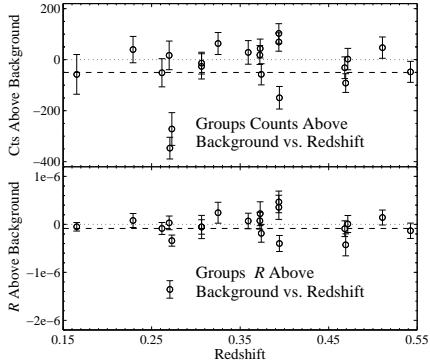}
\caption{Groups intensities above background in D--band (0.2--1.0 keV).
{\it Top Panel}: Counts in groups above estimated local background.
{\it Bottom Pannel}: Surface brightness of groups above background, in units of
$\R$--value -- i.e., counts per square arcsecond per second.
None of the groups is seen more than $5\sigma$ above the background, which is
our minimum criterion for detection.  The dotted line in each panel is the
zero line.  The dashed line in each panel represents the weighted average value
of the quantity on the ordinate of that panel.}
\label{fig:diffplot}
\end{figure}

Ultimately, because of the spatial variability of the background, which is
evident in Figure~\ref{fig:histfig}, the most useful description of the
background is probably the local one.  The top panel of
Figure~\ref{fig:diffplot} shows the number of counts by which each of the 21
groups in our sample is brighter than its estimated local background, plotted
against redshift.  The bottom panel of that figure shows the surface
brightness above the local background $SB_g$, again plotted against redshift.
The 1--$\sigma$ error bars are computed on the ordinates as the summation in
quadrature of the $N^{1/2}$ Poisson noise from the group ($\sigma_g$) and from
the background ($\sigma_{BG}$).  No group is seen more than $5\sigma$ above
the background, which is our minimum criterion for detection.  The dashed line
in each panel represents the
average value of the
quantity on the ordinate of that panel, weighted by the product of the number
of pixels and the effective exposure time.

In those cases where both the X--ray and the optical centers of low luminosity
groups are known, the correlation between the two shows considerable variance
\citep{mulchaey2000}.  Because of the possibility that the X--ray centers
might not coincide with the optical centers of the groups in our survey, we
paid particular attention to the analysis with an aperture of radius 500 kpc
around each group, because an aperture this size should be large enough to
capture a large fraction of the X--rays even if there is an offset of
$\sim 0.5\arcmin$ between the X--ray center and the reported optical center.
Inasmuch as no group had an $\R$--value $5\sigma$ or more above the local
background, the results were substantially the same (no plot shown).
Moreover, a greater fraction of groups had negative $SB$--values than when
the $250 {\rm ~kpc}$ radius apertures were used ($15/21$ vs. $11/21$).

We finally note that the groups might potentially contaminate the local
background estimator.  Using reasonable models for the intensity and the
spatial distribution of emission from groups, described in detail in
\S~\ref{subsec:limits} below, we estimate that $\ge 30\%$ of flux comes out
within our aperture, and $\le 10\%$ of flux comes out in the annulus from
1.05 to 1.20 times the aperture radius.  This portion of group flux that falls
within the annulus that we use to define the local background ought to
increase the local background measurements, and therefore to depress the
surface brightness measurements.  The depression, however, is not severe:
since the group surface brightness profile is expected to peak toward the
center, the surface brightness is still expected to be positive.  Moreover,
the average $\R$--value of the local background samples was
$2.9\times10^{-6}{\rm~cts~arcsec^{-2}~s^{-1}}$, which is only slightly more
than the
modal $\R$--value of the randomly placed apertures, and slightly less than
their median value.  It therefore does not appear that the local background
estimator was significantly increased by flux from the groups.

As a final test of whether the estimate of the background was contaminated by
group--flux, we used an annulus farther from the aperture: the annulus
from 1.5 to 2.0 times the aperture radius.  This change had no important
effect on our results.

\section{Spectral Analysis}
\label{sec:spec}
We also explored whether examining the shape of the X--ray spectrum would
enhance the contrast to the background in the search for groups.
In the simplest possible test, we
investigated whether the spectral energy distribution (SED) of
photons received from the locations of groups differed from that of
photons from elsewhere.  The background consists of the particle
background, and the combined diffuse glows of our galaxy, the
Galactic halo, the local group, and extragalactic point--sources and
diffuse sources \citep{mccammon+sanders1990, mccammon_et_al2002}.

\begin{figure}[t]
\plotone{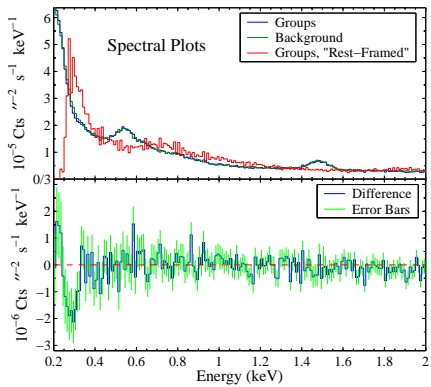}
\caption{Spectra showing $\R$--values per 10 eV bin..  The rest--framed groups
show no signs of any spectral features
{\it Top Panel}: Groups, background, rest--framed groups spectra, 0.2--2 keV.
Spectrum for locations of groups is in blue, for background is in green, and
for rest--framed groups is in red.
{\it Bottom Panel}: Difference spectrum, 0.2--2 keV.  1--$\sigma$ error bars
are in green.}
\label{fig:spec}
\end{figure}

Spectral results from our observations are presented in
Figure~\ref{fig:spec}.  In our data, the spectrum of the groups is
indistinguishable from that of the background.  This can be seen in
several ways.  The top panel of Figure~\ref{fig:spec} show,
with a bin--size of $10 {\rm~eV}$, the stacked spectrum of the groups in
blue, the spectrum of the background in green, and the ``rest--framed''
spectrum of the groups in red (that is, the energy of each photon from the
groups is multiplied by $(1+z_{\rm group})$ prior to binning).  The groups
and the background both show broad emission features centered at $\sim
0.55 {\rm~keV}$ and $\sim 1.5 {\rm~keV}$ (top panel of
Figure~\ref{fig:spec}), that are most likely due to O VII at redshift 0
and neutral Si on the detector, respectively.
The difference spectrum, however, is
0 to within the 1--$\sigma$ error bars (bottom panel), indicating that the SED
from regions of groups is identical to that of other regions.  Furthermore, the
rest--framed groups spectrum shows no signs of any spectral features.

\section{Discussion}
\label{sec:disc}
In recent years, research efforts have increasingly been devoted to
characterizing the extended X--ray emission from collapsed halos of a variety
of masses, from clusters to groups to galaxy--mass halos.  For two examples
(among many) of studies of gas in higher--mass systems -- clusters,
\citet*{akahori+masai2005} investigate the $L_X$--$T_X$ scaling relation
describing intracluster gas, and \citet*{arnaud_et_al2005} investigate the
$M$--$T$ scaling relation of intracluster gas.

Our study is concerned with the properties of gas in lower--mass systems.
\citet{xue+wu2000} attempt to bridge the gap between clusters and groups in
the local universe by performing an analysis of the $L_X$--$T$,
$L_X$--$\sigma_v$, and $\sigma_v$--$T$ relations of 66 groups and 274 clusters
taken from the literature.
The \textsl{XMM}--Large Scale Structure survey identified groups and clusters
out to redshift 0.6 and beyond, and \citet{willis_et_al2005} examine the
$L_X$--$T_X$ relation of the objects in their survey.  Interestingly, in
contrast to the analysis in this paper, they detect X--ray emission from
low--mass objects out to $z=0.558$.
The groups that they found were X--ray selected, so the selection effect
surely contributed to their finding groups in their survey while
we did not in ours (although we note that this does not explain the physical
difference that is responsible for the discrepancy in X--ray luminosities).

\subsection{Fits to $L_X$--$\sigma_v$ Data}
\label{subsec:fits}
In order to estimate how our data ought to have appeared, we need to know
what X--ray luminosity should be expected from a group with a given mass or
velocity dispersion.
The groups of the M03 catalog are a convenient sample of nearby groups with
known velocity dispersion and X--ray luminosity.  Since any fit to the
$L_X$--$\sigma_v$ relation in the M03 groups will show scatter, a fit is
not a line or curve but a 2-dimensional region in $L_X$ versus $\sigma_v$
space.

In fact, properly, one would want to construct a 2D probability
distribution function on $L_X$--$\sigma_v$ space based on the distribution
of the M03 sample.  This function would give the probability density for a
group drawn from the same population as the reference sample to have a
particular velocity dispersion and X--ray luminosity. One could then
properly examine the likelihood that the \textsl{CNOC2} groups
are drawn from the same population as the M03 groups.

In practice, it would be difficult to construct the 2-D distribution function
we require, because the M03 sample is not dense enough in $L_X$--$\sigma_v$
space.  We therefore reverted to a conventional parametric analysis, based on
linear fits to $(\log \sigma_v, \log L_X)$ data.  We will present several
linear fits to the M03 data, and we describe in detail our choice for
which linear fit is most appropriate.  

PT04 examine the relationship between the X--ray
luminosity ($L_X$) and the velocity dispersion ($\sigma_v$) of galaxy groups in
the M03 catalog.  They perform regression analysis on
$\log L_X$ and $\log \sigma_v$ and find that, as with any regression in
which there is significant scatter, which line is considered the best--fit
depends on which variable is considered to be independent and which dependent.
We repeated their analysis of the M03 groups and obtained best--fit lines that
are very similar (though not identical) to those that they obtained.  In
this paper, we present our own fits to the data.
If velocity dispersion is taken to be the independent variable, the best--fit
regression line is
\begin{equation}
\label{eq:M03D}
\log \left(L_X / 1 {\rm~erg~s^{-1}}\right) = 36.8 + 2.08 \log \left( \sigma_v / 1 {\rm~km~s^{-1}} \right).
\end{equation}
If, on the other hand, luminosity is taken to be the independent variable,
the best--fit regression line (what PT04 refer to as the ``inverse regression
line'') is
\begin{equation}
\label{eq:M03I}
\log \left(L_X / 1 {\rm~erg~s^{-1}}\right) = 30.9 + 4.50 \log \left( \sigma_v / 1 {\rm~km~s^{-1}} \right).
\end{equation}
Data from M03 and best--fit lines are displayed in Figure~\ref{fig:maxlums}.

Which regression represents the ``true'' relation between the X--ray
luminosity and the velocity dispersion of groups of galaxies?
In Appendix~I, we present a more detailed analysis of the relative merits of
various procedures for fitting lines to data.  The result of this analysis
is that neither direct nor inverse regression is ideal,
because surely the measurements of both $\sigma_v$ and $L_X$
were subject to errors, and furthermore there is
undoubtedly intrinsic scatter in the $L_X$--$\sigma_v$ relation.  The better
relation out of these two is the one in which the variable taken to be
independent has
lower relative uncertainty.
In our case, this is probably the inverse regression,
because a velocity dispersion determined from only a few galaxies is
likely to be quite uncertain.  Furthermore, the steeper (inverse) relation
comes closer to fitting the high--$\sigma_v$, high--$L_X$ end of the
distribution of groups, which is the end that we are most interested in
when looking for groups at $z\sim 0.5$.
Still, although the inverse regression is probably preferable to the direct
regression, since there is scatter in both
variables, this situation calls for least squares orthogonal distance
fitting (described in Appendix~I).  If we assume that the typical relative
errors on $\log(\sigma_v)$ and
$\log(L_X)$ are the same (i.e., the error is shared equally between the two
variables), then the distance--fit line,
\begin{equation}
\log \left(L_X / 1 {\rm~erg~s^{-1}}\right) = 31.5 + 4.25 \log \left( \sigma_v / 1 {\rm~km~s^{-1}} \right),
\label{eq:distfit}
\end{equation}
is nearly
as steep as the inverse regression line, as shown in Figure~\ref{fig:maxlums}.

\subsection{Limits on Group Luminosities}
\label{subsec:limits}
Since none of the groups was visible at $5\sigma$ above the background,
we may place limits on their fluxes and therefore, since we know their
redshifts, on their luminosities.
The total number of background counts $TC_{BG}$ at the location of a group is
estimated from the smoothed background image.  Assuming Poisson statistics,
the standard deviation on this number is
$\sigma_{TC_{BG}} = (TC_{BG})^{1/2}$.  The maximum number of group counts $TC_G$,
then, is limited by
\begin{equation}
TC_G < 5\sigma_{TC_{BG}} = 5 \sqrt{TC_{BG}},
\label{eq:TCG}
\end{equation}
We estimate each group's temperature with the standard relation given in,
e.g., \citet{mulchaey2000}:
\begin{equation}
T = \frac{m_p}{\mu} \sigma_v^2 \times (k \beta),
\label{eq:temp}
\end{equation}
where $m_p$ is the proton mass, $\mu$ is the mean molecular weight, $k$ is
Boltzmann's constant, and $\beta$, whose empirical value tends to be around
$2/3$, is the eponymous parameter of the isothermal--$\beta$ model for 
density and may be considered to be defined by this equation.  (For low-density
objects, $\beta$ has been measured to be closer to $1$, but for analytical
simplicity we used $\beta=2/3$ for most calculations.)
The $\beta$ density model for a spherical isothermal plasma is
\begin{equation}
\rho(r) = \rho_0 \left(1 + \left(\frac{r}{r_c}\right)^2\right)^{-3\beta/2},
\label{eq:beta}
\end{equation}
where $\rho_0$ is the central density, $r$ is the radial distance from the
sphere's center, and $r_c$ is the ``core radius,'' a distance within which
the density is nearly constant (approximately $\rho_0$).
Using the \textsl{Astrophysical Plasma Emission Code} (\textsl{APEC}) model
in \textsl{XSPEC}, we solve for what X--ray luminosity, at a given
temperature, would lead to the number of group counts given by the maximum
possible value of $TC_G$ in (\ref{eq:TCG}).

\begin{figure}[t]
\plotone{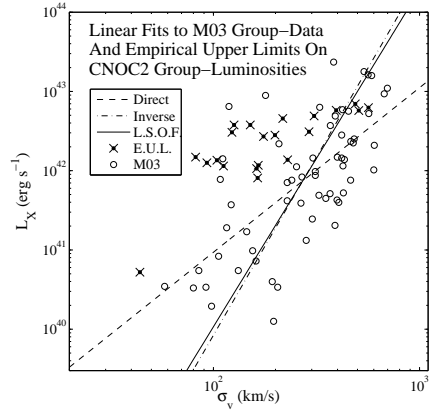}
\caption{Linear Fits to M03 Group--Data And Empirical Upper Limits On CNOC2
Group--Luminosities: bolometric X--ray luminosity vs. line--of--sight velocity
dispersion.
The circles show the M03 groups. The dashed line is the direct regression line,
the dashed--dotted line is 
the inverse regression line, and the solid line (L.S.O.F.) is the 
least squares orthogonal fit line.
These three lines are fits to the M03 data.
The X's (E.U.L.) display the ``empirical upper limits'' to groups' luminosities,
i.e., the maximum luminosities that the \textsl{CNOC2} groups could
have had during our observations without showing up at more than $5\sigma$
above the background.
The uncertainty in velocity dispersion associated with the X's is typically
nearly $\sim100\%$, as shown in Table~\ref{ta:21Groups}.
Three of the X's lie beneath both the dashed--dotted and the solid lines,
indicating that these three groups were predicted to be luminous enough to be
detected, and yet were not.
}
\label{fig:maxlums}
\end{figure}

In Figure~\ref{fig:maxlums}, the X's show the maximum luminosity that
each group could have had during our observations
(if the group were more luminous, it would have
shown up at $>5\sigma$ above the background).
If a group is predicted to be less luminous than the luminosity represented by
its X, then it is no surprise that we failed to see the group; conversely, if
a group is predicted to be more luminous than the luminosity represented by
its X, then the group is less luminous than predicted.
The dashed line shows the $L_X$--$\sigma_v$
direct regression line, the dashed--dotted line shows the inverse regression
line, and the solid line shows the distance--fit line.
All X's (all groups) lie above the direct regression line, so \emph{if} that
line represents the true relationship between the variables, there is no
surprise in seeing none of the groups.  Three groups, however, lie below both
the inverse regression line and the distance--fit line; so if those lines
represent the true relationship between the variables, it might be surprising
that we failed to see three of the groups.  In the remainder of this section,
we will try to quantify how surprising it is.

\subsection{Quantifying Surprise}
\label{subsec:surprise}
The \textsl{CNOC2} catalog of small groups \citep{carlberg2001gg}
contains groups with as few as three galaxies in them.  Any measurement of the
velocity dispersion of such a group will necessarily have large
uncertainties.  In fact, for some groups, the estimated uncertainty in velocity
dispersion is actually greater than the velocity dispersion itself.  Because
the symbol $\sigma$ is frequently used to denote both uncertainty and velocity
dispersion, it is inelegant to represent the uncertainty in velocity
dispersion, but, eschewing elegance, we shall use $\sigma_{\sigma_v}$ to
denote uncertainty on $\sigma_v$.

It is possible that our null result -- our failure to detect any of the
groups in our survey -- occurred because the velocity dispersions of the three
groups predicted to be observed were overestimated.  If so, then perhaps their
predicted luminosities based on, e.g., (\ref{eq:distfit}), should have been
low enough that the groups should not have been seen after all.
Of course, if there were no bias to the measurements of $\sigma_v$,
it is just as likely that the velocity dispersions of these three groups were
{\it underestimated}, which would make their non--detection even more
surprising.  Furthermore, in order for this sort of statistical error to
explain our null result, not only would the velocity dispersions of the
three groups that we think we ought to have detected need to have been
sufficiently overestimated, but no {\it other} groups could have have had their
velocity dispersions underestimated by too much.  In short, the statistical
errors would need to be in particular directions on particular groups.  If
there is no non--statistical reason for measurements to be in error, then what
is the likelihood that the types of error needed to explain our null result
actually occurred?

To answer this question, we simulated 10,000 mock--\textsl{CNOC2} catalogs,
each with its own set of mock--velocity dispersions
($\{\widehat{\sigma_v}(i)\}$) associated with the mock--groups.  If the set of
measured velocity dispersions from the actual \textsl{CNOC2} catalog
is $\{\sigma_v(i)\}$, then, for each mock--catalog, the simulated velocity
dispersion of the $i$th group $\widehat{\sigma_v}(i)$ was drawn from from a
gamma distribution with mean $\sigma_v(i)$ and standard deviation
$\sigma_{\sigma_v}(i)$.  The gamma distribution has the attractive property
that the mean and the standard deviation may be independently specified.  
Because the tail of the gamma distribution extends to infinity, however, a small
fraction of mock--groups were initially given unrealistically high velocity
dispersions. Incidentally, this is preferable to the normal distribution, which
has another tail that extends below zero to negative infinity -- unphysical
values for velocity dispersion.
To deal with the gamma distribution's problem of predicting a few
unrealistically high values, we, for each group $i$, eliminated from
the set of mock--catalogs all mock--groups with velocity dispersions more than
$4\sigma_{\sigma_v}(i)$ above the mean ($\sigma_v(i)$), and replaced these values
with new gamma--distributed variables.  We repeated this procedure iteratively
until there were no more mock--groups with velocity dispersion more than
$4\sigma_{\sigma_v}$ above the mean.  Because this procedure involves replacing
the high--end outliers with values closer to the mean, it had the effect of
slightly depressing the mean.  All values were therefore increased by the
difference between the new mean and $\sigma_v(i)$, thereby raising the mean
back to $\sigma_v(i)$, as it should be.

We also tried several other random distributions for assigning mock--velocity
dispersions to our mock groups.  We tried a gamma distribution with no high--end
cut--off; and we tried several modifications of the normal distribution,
all of which included a low--end cut--off at $0$.  We found that those results
were not importantly different from results from the gamma distribution
with the high--end cut--off, which is the analytically simplest distribution
that produces realistic results.  Results are presented only for the
mock--catalogs generated with the high--end cut--off gamma distribution.

Each mock--group was assigned the same aperture size as the corresponding
real group.  Its mock--intensity was spatially distributed according to a
spherical isothermal--$\beta$ model.  Since emissivity is proportional to
the square of density $\rho$, the surface brightness or intensity $I$ is
proportional to the integral along the line--of--sight of $\rho^2$.  At a 
projected viewing ``impact parameter'' $b$ from the sphere's center,
\[
I(b) \propto \int_{r=b}^{r=r_h} \rho(r)^2 {\rm d} s,
\]
where $r_h$, the ``halo radius,'' is the outer boundary of the sphere of
plasma, and $s$ is the variable along the line--of--sight.  Substituting for
$\rho$ with equation~(\ref{eq:beta}) and performing the integration results in
the following expression for the surface brightness:
\begin{equation}
\mathcal{B}(x,A,\beta) \propto (1+x^2)^{-3\beta} (A^2 - x^2)^{1/2} \spa { _2F_1}\left(\frac{1}{2},3\beta,\frac{3}{2},\frac{x^2-A^2}{1+x^2} \right),
\label{eq:bright}
\end{equation}
where $x \equiv b/r_c$ and $A \equiv r_h/r_c$ are substitutions to
simplify the integration, and $_2F_1$ is the Gauss hypergeometric function.
Three quantities go into the normalization of $\mathcal{B}$ for each
mock--group: its mock--luminosity from e.g. (\ref{eq:distfit}); its luminosity
distance $D_L(z)$; and its core radius.  Finally, we assumed a foreground
column density of neutral hydrogen of $2\times10^{20}{\rm ~cm^{-2}}$
\citep{dicke+lockman1990}.

It was unclear what core radii our mock--groups should be assigned, because
the $\beta$-model core radii of groups and clusters vary a great deal.
\citet*{ota+mitsuda2004} report that in a sample of 79 clusters, including 45
``regular'' clusters and 34 ``irregular'' clusters, the mean of all 79 values
of $r_c$ is $(163 \pm 202)~{h_{70}}^{-1}{\rm~kpc}$, midway between the mean
$r_c$ of the regular clusters ($(76 \pm 60)~{h_{70}}^{-1}{\rm~kpc}$) and the
mean $r_c$ of the irregular clusters ($(273 \pm 259)~{h_{70}}^{-1}{\rm~kpc}$).
Because there is no strong correlation between $r_c$ and group--mass, we think
it is reasonable to adopt a uniform assumed value of $r_c$ for the mock--groups
in our mock--catalogs.  We ultimately decided to set each mock--group's
core radius to $125 {\rm~kpc}$, because this value lies in the middle
of the range of values reported by \citet*{ota+mitsuda2004}.  So long as the
value of $r_c$ that we adopted is physically plausible, the precise value
does not matter much, because the predicted number of counts from a
group, within an aperture of fixed size, is not very sensitive to the assumed
value of $r_c$, as long as the angular--size of the aperture is larger than
that of the core radius.

The SED of each mock--group (and hence the flux
in a given energy band) is calculated with \textsl{XSPEC},
using \textsl{APEC}, with its mock--temperature calculated from
equation~(\ref{eq:temp}).  For $A$, we tried values between $2$ and $\infty$,
and for $\beta$, we tried values between $0.6$ and $1.0$.  The results
presented below are for $A=\infty$ and $\beta=2/3$, and for D--band
(0.2--1.0~keV).

\begin{figure}[t]
\plotone{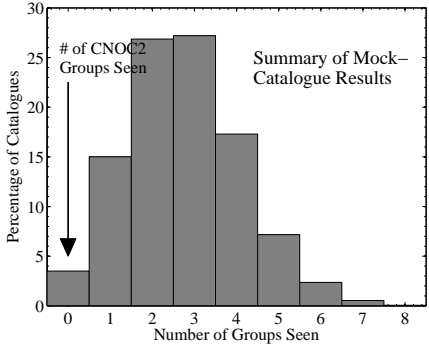}
\caption{Histogram of the results of the mock--catalog analysis.  Of the
10,000 mock--catalogs, $96.5\%$ contained at least 1 mock--group visible at
the $5\sigma$ level.  In the actual \textsl{CNOC2} survey no groups were
visible at this level.
}
\label{fig:histsim}
\end{figure}

In brief, the results of this Monte Carlo indicate that the probability
appears to be small, but not vanishingly so, that statistical fluctuations in
the measured values of $\sigma_v$ explain our null result.
Figure~\ref{fig:histsim} shows a summary of the mock--catalog results.
The median and the mode number of groups seen per catalog are both $3$, and
the mean number seen is $3.1$, which is in good agreement with the prediction of
section~\ref{subsec:limits}.  The finding of our observation -- that no
groups were seen -- occurred in only $3.5\%$ of the mock--catalogs.
If there were no systematic errors in the measurements of velocity
dispersions of the \textsl{CNOC2} groups, then it is unlikely but not
impossible that the only reason we had expected some of the groups to be
visible in the first place was erroneously high measurements of $\sigma_v$.

\vspace{0.5 in}
\subsection{Redshifted Mulchaey Groups}
\label{subsec:M03z}
Perhaps we have placed too much emphasis, in sections \ref{subsec:fits} and
\ref{subsec:limits}, on which best--fit line actually fits
the M03 data the best -- when the truth is that there will be significant
scatter to the data around any line drawn through the cloud of points.
An alternative way to determine whether it is surprising that no groups in
our survey showed up at $5\sigma$ above the background is to ask
whether the \textsl{CNOC2} groups and the M03 groups were drawn from the same
population.  We addressed this question by investigating how many of
the groups in the M03 atlas would have been visible at the $5\sigma$ level if
they had been in our observation.

The M03 groups are all in the local universe, approximately at redshift 0.
If they had been in our
survey, i.e., between redshifts 0.1 and 0.6, they would have both been dimmer
and subtended a smaller solid angle on the sky.  Furthermore, they would have
been observed with \textsl{XMM--Newton}, 
which, owing to its greater focal ratio, has a higher particle background
rate than \textsl{ROSAT} (the instrument with which they were originally
observed).

There were 109 groups in the M03 catalog, of which 61 were seen in X--rays.
We simulated observations of each of these 61 M03 groups at each of the 21
redshifts of the \textsl{CNOC2} groups,
ranging from $z_1 = 0.165$ to $z_{21} = 0.543$ (ID numbers taken from
Table~\ref{ta:21Groups}).  As before, our criterion
was that a group must outshine the background by at least $5\sigma$ to qualify
as being detected in our simulated observations.  Each group was assumed to be
at a location with detector--averaged characteristics: the mean exposure map
value, the median background value (from Figure~\ref{fig:histsim}), the mean
amount of smearing due to the point--spread function, and the mean amount of
lost usable detector--area because of bright point sources.
Much like in the mock--catalog calculations described in
section \ref{subsec:surprise}, we calculated the count--rate of each group
with \textsl{XSPEC}, using \textsl{APEC}, with the group's temperature
calculated from equation~(\ref{eq:temp}).  We distributed emission according to
the isothermal--$\beta$ profile using equation~(\ref{eq:bright}), with
$\beta=2/3$, $A = \infty$,
and $r_c = 125 {\rm~kpc}$.  For our simulated observations, we used an aperture
of $250 {\rm~kpc}$, and looked in D--band.  For background, we used
$\R_{BG} = 3.0\times10^{-6} {\rm~cts~arcsec^{-2}~s^{-1}}$.  Finally, as before,
we assumed a foreground
column density of neutral hydrogen of $2\times10^{20}{\rm ~cm^{-2}}$.
Within our simulation, the number of M03 groups that exceeded the $5\sigma$
detection criterion varied from 22 of the 61 total groups (or $p_1 = 36\%$)
at the lowest redshift ($z_1$), to 2 of 61 ($p_{21} = 3\%$) at the highest
redshift ($z_{21}$).

We identify two populations of the Mulchaey groups -- all 109, and the 61 that
were seen in X--rays.  In Appendix~II, we present in detail our estimate of the
likelihood that the \textsl{CNOC2} groups and the M03 groups were drawn from
the same population.  The result is that the probability of obtaining our null
result on the assumption that the \textsl{CNOC2} groups were drawn from the
same population as the whole M03 catalog is 14.7\%; and the
analogous probability on the assumption that the \textsl{CNOC2} groups were
drawn from the same population as the 61 X--ray--bright M03 groups is 2.8\%.
These probabilities (especially the latter) are low enough to be interesting,
but are certainly not dispositive.

\section{Conclusion}
\label{sec:conc}
The analysis in section \ref{subsec:M03z} is mildly suggestive that the
groups of the \textsl{CNOC2} survey constitute a sample that is on average less
luminous than the sample of groups observed by Mulchaey and collaborators in
the 2003 atlas.  The interesting comparison, however, is not between two
particular samples, but rather between the population of groups at intermediate
redshift and the population in the local universe.

A first caveat is that the two samples were selected in very different ways,
and there is no {\it a priori} reason to suspect that their underlying
populations have similar luminosity functions.
Furthermore, it is conceivable that for some reason the M03 groups should be
expected to be a more luminous sample of their population than the
\textsl{CNOC2} groups are of theirs.  Perhaps the M03 groups are on average
more massive (although the ranges and distributions of velocity dispersions for
the two samples are nearly identical).  Do the M03 groups contain more galaxies
than the \textsl{CNOC2} groups?  Certainly, Mulchaey reports more galaxies per
group on average -- including as many as 63 galaxies in one group.  Of the
\textsl{CNOC2} groups in our field of view, none contained more than 4
galaxies with identified redshifts.  But most or all of the difference here is
almost certainly due to the difficulty of measuring spectroscopic redshifts of
faint galaxies at large distances; we therefore see no reason to think that
the M03 groups are on average more massive than the \textsl{CNOC2} groups.

An obvious criticism of the present work is that the \textsl{CNOC2} groups
might not be groups at all, but just chance superpositions of galaxies.
While this would clearly explain the lack of X--ray emission, we will argue
that we find it unlikely to be the reason why we detected no hot gas.

\citet{carlberg2001gg} identified as a group any object in the
\textsl{CNOC2} Survey that
contains at least 3 galaxies in sufficiently close redshift--space proximity.
This identification procedure does leave open the possibility of confusion:
if two galaxies are separated by a large distance, but have appropriately large
peculiar velocities, they could have similar recessional velocities (redshifts)
despite not being part of a gravitationally bound structure.  Or, even if a
few galaxies are close to one another in physical space, they will not be
part of a bound system without a dark matter halo that is massive 
enough to bind them.

In searching for groups and clusters, it is often assumed that where there
are associations of galaxies, there is dark matter to bind them, but this
assumption could be wrong.  With rich clusters, there are multiple independent
checks of this assumption: the temperature of intergalactic gas (which can be
detected through its X--ray emission or through inverse Compton distortions
to the cosmic microwave background) is one indicator of the mass; the
velocity--distribution of galaxies is another indicator of the mass, and
in particular the distribution should be Maxwellian (or Gaussian) for a
virialized structure;
and the degree of gravitational lensing of background galaxies is a third
indicator of the mass.

The first check has obviously failed in this case -- for these
21 groups, there is no apparent hot gas component.  The second check is
useless for an association of very few (3--6) galaxies -- nearly any velocity
distribution is consistent with an underlying Gaussian distribution.  How about
lensing?  For structures of $\lsim 10^{13.5} M_\sun$, noise prevents
detection of a lensing signal.
As a result, in the \textsl{CNOC2} fields, it is only possible to determine a
statistical lensing mass for the groups by stacking them and determining their
combined lensing signal.
This analysis has been performed by
\citet{hoekstra_et_al2001, hoekstra_et_al2003CNOC2}:
their calculated estimate of the average
velocity dispersion per group based on the mass inferred from the average
shearing of background galaxies ($258_{-56}^{+45} {\rm~km~s^{-1}}$ for
$\Omega_m=0.2$ and $\Omega_\Lambda=0.8$) is in good agreement with the average
velocity dispersion of $230 {\rm~km~s^{-1}}$ from the kinematics of the groups'
member--galaxies.
The observed lensing signal was sufficiently strong that chance projections of
unbound galaxies alone would have difficulty mimicking it.
Some of the detected lensing mass is therefore almost surely in the form of
galaxies orbiting within a parent dark matter halo -- groups.  It is entirely
reasonable to expect part of this group--component of the lensing mass to
consist of hot gas trapped within the group potential.
On the other hand, it is not unlikely that a portion of the lensing mass
is in the form of projected, yet physically unassociated, galaxies.  Although
there may be limited amounts of gas associated with the individual galaxies,
such projections will produce no extended group emission.
Therefore, the presence of a lensing signal does not automatically lead to
the expectation of group--like X--ray emission that is at sufficient
surface--brightness to be detectable in our survey.
%

It is conceivable that there were several relatively massive groups, or
dark clusters (i.e., cluster--mass objects with very low mass--to--light
ratios), that dominated the mean mass measurement.
Although we cannot rule out the possibility that a large fraction of the mass
is in only a few objects that happened to fall outside the field of view of our
observations, and that the remaining objects are not gravitationally bound
structures but rather chance associations of galaxies that just happen to be
near one another in redshift--space, 
this seems somewhat unlikely:
Combinatorial statistics dictate that if there were a few
massive halos scattered among the 40 alleged groups and none of the rest are
true groups, then the probability is low that all the massive ones ended up
among the 19 groups that did not fit into our field of view.  
Consider, for instance, if there were only four massive halos:
since 19 is roughly half of 40, if the four massive halos were randomly
distributed the probability that all four would be out of our field of view is
$\lsim\left(1/2\right)^4$.
Furthermore, a calculation in \citet{carlberg2001gg} shows that the number of
alleged groups in the field in the redshift range $0.1<z<0.6$
(40 in the \textsl{CNOC2} field, 21 in the \textsl{EPIC--PN} field of view)
is roughly consistent with predictions.
The concordances both of the observed number of groups with the predicted
number, and of the observed velocity dispersions with the mass inferred from
weak lensing, together constitute a strong indication that the \textsl{CNOC2}
group sample does not suffer significant contamination from chance
superpositions of galaxies.
An important caveat is that the number of groups in the \textsl{CNOC2} fields
in the high $\sigma_v$ range is in excess of the aforementioned predictions
derived from the basic formalism of \citet{press+schecter1974} by a factor of
$\sim 4$.  Based on the \citet{carlberg2001eg} results, $\sim 80\%$ of such
high $\sigma_v$ groups could be false, and our own null results are consistent
with such this hypothesis.

If we take seriously the conclusion, for which our analysis has generated some
modest evidence, that galaxy groups at intermediate
redshift are less X--ray luminous than those at redshift zero, we can think of
several potential explanations.  Groups at $z \sim 0.5$ are younger than their
relatives in the local universe, which leads to the following three
possibilities: \pn{i} Since simulations and observations 
indicate that groups
and clusters at the present are accreting gas from cosmic filaments, groups at
a younger stage of the universe would presumably have been accreting gas for
less time than the more evolved groups of the present--day universe and might
therefore contain less gas.  \pn{ii} Even if the intragroup gas in young groups
is not lower density than the gas in groups today, if the young groups are still
in the process of heating up --
if they have not yet reached their virial temperatures -- their luminosities
would be reduced. (We find this possibility to be unlikely, because the
heating timescale is short relative to a Hubble time, even at $z \sim 0.5$.)
\pn{iii} Even if the gas in young groups is on average the same density and
the same temperature as the gas in nearby groups, if it is of lower
metallicity its X--ray emissivity would be reduced.  There are undoubtedly
feedback processes that enrich the intragroup medium, and pristine gas of
primordial abundance is also surely constantly falling onto groups, so it is
not clear whether to expect higher metallicity in the IGM of groups that are
at $z \sim 0.5$ or at $z \sim 0$. For warm gas, observed in a soft X--ray band
such as our D--band, line--emission can be important, depending on abundances.
For example, if a $T = 1{\rm~keV}$ group at redshift 0.5 contained primordial
gas, its D--band X--ray flux would be only two thirds as great
as if its gas were
at 10\% solar abundance and only half great
as if its gas were at 20\% solar
abundance.  It is possible, therefore, that the relation between velocity
dispersion and X--ray luminosity ought to be shifted down by a factor of
$\sim 2$ in Figure~\ref{fig:maxlums}.  Such a shift would make it less
surprising that we detected no X--ray emission from the groups.

Future data on high--redshift clusters that make it possible to examine the
time--evolution of the density and metallicity of the intracluster medium
in low--mass clusters, and deeper searches for X--ray gas in low--mass objects
at intermediate redshift, will help to answer why our survey failed to
detect such gas.

\vspace{\baselineskip}
\vspace{\baselineskip}
We thank Kevin Roy Briggs and Maurice Leutenegger for invaluable advice on
reduction and analysis of \textsl{XMM--Newton} data;
and we thank Fred Blumberg for extremely helpful
comments and criticism.  We furthermore would like to acknowledge the many
incisive observations and constructive suggestions from our referee and from
our scientific editor.
FP acknowledges support by NASA through grant NAG5-13354, and
CAS acknowledges support by NASA grant SAOG03-4158A.
This research has made use of publicly available data from the
\textsl{XMM-Newton} Science Archive.

\section*{Appendix I}
\label{appendixI}

In this Appendix, we address the question of which regression represents the
``true'' relation between the velocity dispersion and the X--ray luminosity of
groups of galaxies.  It is instructive first to consider a statistical
effect that is common to all regression analysis.  It has been known at
least since Karl Pearson's seminal paper in 1901 that if there is scatter (say,
due to measurement error) in the variable considered to be the independent one,
this will in general cause the recovered slope from regression analysis to
be shallower than the slope of the true underlying relation between the
variables.  (PT04 demonstrate a nice example of this effect.) An
important consequence of this effect is that when there is scatter in both
variables, the slopes from both the direct and the inverse regression will
be shallower than the respective true slopes. 

For example, suppose two variables $x$ and $y$ are related by
$y = \alpha x$ (or $x = \beta y$, where $\beta = 1/\alpha$).  In an experiment,
a large number of $(x,y)$ pairs are measured, and there are statistical
uncertainties in the measurements of both variables.  If $x$ is considered
to be the independent variable, regression analysis will lead to
$y = \hat{\alpha} x$, where in general $\hat{\alpha} < \alpha$
because of the scatter in $x$.  If $y$ is considered to be the
independent variable, regression (``inverse regression'') analysis
will lead to $x = \tilde{\beta} y$, where in general $\tilde{\beta} < \beta$
because of the scatter in $y$.  Of course, if we wish to represent $y$ as
a function of $x$ even in the case of the inverse regression, we will write
$y = \tilde{\alpha} x$, where $\tilde{\alpha} = 1/\tilde{\beta} > \alpha$.
So, although the recovered slopes are in general less than the respective true
slopes for both regressions
($\hat{\alpha} < \alpha$ and $\tilde{\beta} < \beta$),
when we fix the abscissa and ordinate variables we get a shallower slope
than the true one in the case of the direct regression and and a steeper
slope in the case of the inverse regression.  Note that this effect holds
whether the scatter comes from measurement error or is intrinsic to the
variables.

When there is scatter in both variables, so that neither one is a proper
independent variable, it is sometimes appropriate to use a type
of best--fit line that depends on the geometric orientation of the points.
Instead of finding the line that minimizes the summed, squared {\it vertical}
deviations of the data from the line, as in the case of regression,
some situations are better analyzed by finding the line that minimizes the
summed, squared {\it perpendicular} distances of the data from the line.  This
is sometimes called ``least squares orthogonal distance fitting,'' and was the
subject of \citet{pearson1901}.  Because this type of fitting,
in contrast to linear regression, is unfortunately sensitive to the units of
measurement, it is advisable to normalize the data prior to fitting.  The most
common normalization is by the estimated errors or uncertainties ($\sigma_x$
and $\sigma_y$) on the variables \citep{NR1992, akritas+bershady1996}.
The direct regression line will have the minimum slope, the inverse regression
line will have the greatest slope, and the ``distance--fit'' line will have an
intermediate slope (closer to the direct regression slope if the scatter in
$x$ is smaller; closer to the inverse regression slope otherwise).  All three
lines intersect at the centroid of the data.

For completeness, we present the equation of the least squares orthogonal
distance fit line.  Let us express the measured data as a set of $N$ points
$\{(x_i, y_i)\}_{i=1}^N$.
To simplify the form of the line's equation, we define the following sums
of data:
\[
S_x \equiv \sum_{i=1}^N x_i \qquad S_y \equiv \sum_{i=1}^N y_i
\]
\begin{equation}
\label{eq:sums} S_{x^2} \equiv \sum_{i=1}^N {x_i}^2
 \quad S_{xy}  \equiv \sum_{i=1}^N (x_i \: y_i)
 \quad S_{y^2} \equiv \sum_{i=1}^N {y_i}^2;
\end{equation}
and we define the following combinations of these sums:
\begin{eqnarray}
\nonumber      A & \equiv & S_x S_y - NS_{xy}\\
\nonumber      B & \equiv & (S_x)^2 - (S_y)^2 - N(S_{x^2}-S_{y^2})\\
\label{eq:ABC} C & \equiv & - S_x S_y + NS_{xy}.
\end{eqnarray}
If the equation of the least squares orthogonal distance fit line is
\begin{equation}
y = m x + b,
\label{eq:basiclineeq}
\end{equation}
then the slope of the line is given by
\begin{equation}
m = \frac{-B \pm \sqrt{B^2 - 4AC}}{2A},
\label{eq:meq}
\end{equation}
and the $y$--intercept is given by
\begin{equation}
b = \frac{S_y - m S_x}{N}.
\label{eq:beq}
\end{equation}
The expression for $m$ in equation~(\ref{eq:meq}) contains a $\pm$--sign from
the solution to a quadratic equation. One solution for $m$ gives the line that
minimizes the summed squared orthogonal distances to the points, and the other
solution gives the line passing through the centroid of the data that maximizes
the summed squared orthogonal distances to the points.

\section*{Appendix II}
\label{appendixII}

In this Appendix, we address how to compute the probability that the
\textsl{CNOC2} groups and the Mulchaey (M03) groups derive from the same
population.  In order to do this, we find the properties of the hypothetical
parent group population $\mathcal{P}$ that maximize the joint probability of
our results and Mulchaey's results.

Let the redshifts of the \textsl{CNOC2} groups be labeled $\{z_k\}_{k=1}^{21}$.
At each redshift $z_k$, let $N_k$ be the number of M03 groups that our
simulation predicts would have been visible in our 110 ksec observation, and
let $p_k$ be the probability that a group drawn from $\mathcal{P}$ will be
visible in our survey if it is in our field of view and at redshift $z_k$.

At redshift $z_k$, the probability of obtaining our result -- of not seeing any
groups -- is
\begin{equation}
q_k \equiv (1-p_k);
\label{eq:qk}
\end{equation}
and the probability of obtaining the M03 result -- of seeing $N_k$ groups, as
predicted by our simulation -- is
\begin{equation}
B(M,N_k,p_k) = {M \choose N_k} {p_k}^{N_k} {q_k}^{M-N_k},
\label{eq:M03prob}
\end{equation}
where $M$ is the total number of groups in the M03 sample, either 109 or 61, as
described above.

The joint probability, then, of obtaining our result and of obtaining the M03
result is
\begin{equation}
P_{\rm joint}(z_k) = q_k B(M,N,p_k) = {M \choose N_k} {p_k}^{N_k} {q_k}^{M-N_k+1}.
\label{eq:joint}
\end{equation}

At each redshift $z_k$, we find the binomial probability $\hat{p}_k$ that
maximizes $P_{\rm joint}(z_k)$.  This set of optimal probabilities
$\{ \hat{p}_k \}$ can be thought of as defining the parent population
$\widehat{\mathcal{P}}$ of groups that maximizes the probability that we would
obtain our null result and that the M03 groups would have their observed
properties.  Each percentage $\hat{p}_k$ in turn can be thought of as an
estimate of the probability that a group would be visible in our survey if it
were randomly selected from $\widehat{\mathcal{P}}$ and placed at redshift
$z_k$.  The value $\hat{q}_k \equiv (1-\hat{p}_k)$, then, is the
probability that such a group would {\it not} be detected in our survey.

If 21 groups were selected at random from $\widehat{\mathcal{P}}$ and placed at
the 21 redshifts $z_k$,
the probability that none would be seen in our survey is the product
of the 21 quantities $\hat{q}_k$.  A maximal estimate of the conditional
probability that none of the \textsl{CNOC2} groups would be detected at the
$5\sigma$ level, given that the \textsl{CNOC2} groups were drawn from the same
population as the M03 groups, may therefore be represented as follows:
\begin{equation}
P(\varnothing ~| {\rm ~same ~pop.}) \leq \prod_{k=1}^{21} \left( 1-\hat{p}_k \right),
\label{eq:condprob}
\end{equation}
where $\varnothing$ denotes our null result of not detecting any
\textsl{CNOC2} groups.  The result of our simulation is that the product in
equation~(\ref{eq:condprob})
is $P \leq 14.7\%$ for $M=109$ and $P \leq 2.8\%$ for $M=61$; this indicates
that the probability is small, but not vanishing, that the \textsl{CNOC2}
groups in our survey have the same luminosity function as the M03 groups.

\bibliography{biblio}

\end{document}